\documentclass[prl,twocolumn]{revtex4}
\usepackage{graphicx}
\usepackage{hyperref}
\usepackage[section]{placeins}
\usepackage{amsmath}
\usepackage{mathtools}
\usepackage{amssymb}
\usepackage{dsfont}
\usepackage{bbold}
\usepackage{ulem}
\newcommand{\ket}[1]{|#1\rangle}             
\newcommand{\braket}[2]{\langle#1|#2\rangle} 
\newcommand{\proj}[1]{| #1 \rangle \langle #1 |}            

\begin{document}

\title{Discriminating single-photon states unambiguously in high dimensions}

\author{Megan Agnew,$^{1,2}$ Eliot Bolduc,$^{2}$ Kevin J. Resch,$^{1}$  Sonja Franke-Arnold,$^{3}$ Jonathan Leach$^{2}$}
\affiliation{$^1$Department of Physics \& Astronomy and Institute for Quantum Computing, University of Waterloo, Waterloo, Canada N2L 3G1}
\affiliation{$^2$SUPA, School of Engineering and Physical Sciences, Heriot-Watt University, Edinburgh EH14 4AS, UK}
\affiliation{$^3$SUPA, School of Physics and Astronomy, University of Glasgow, Glasgow G12 8QQ, UK}
\date{\today}

\begin{abstract}

The ability to uniquely identify a quantum state is integral to quantum science, but for non-orthogonal states, quantum mechanics precludes deterministic, error-free discrimination. However, using the non-deterministic protocol of unambiguous state discrimination (USD) enables error-free differentiation of states, at the cost of a lower frequency of success. We discriminate experimentally between non-orthogonal, high-dimensional states encoded in single photons; our results range from dimension $d=2$ to $d=14$. We quantify the performance of our method by comparing the total measured error rate to the theoretical rate predicted by minimum-error state discrimination. For the chosen states, we find a lower error rate by more than one standard deviation for dimensions up to $d=12$. This method will find immediate application in high-dimensional implementations of quantum information protocols, such as quantum cryptography.

\end{abstract}
\maketitle

Discriminating between different quantum states without error is a fundamental requirement of quantum information science.  However, due to the nature of quantum mechanics, only orthogonal states can be exactly discriminated without error 100\% of the time. In contrast, the discrimination of non-orthogonal states requires a decrease in either detection accuracy, using minimum-error state discrimination, or detection frequency, using unambiguous state discrimination. Minimum-error state discrimination (MESD) always provides information about the state, though the information may be incorrect \cite{Helstrom}. Conversely, unambiguous state discrimination (USD) provides either the correct information about a detected state or inconclusive information about the state \cite{Ivanovic1987,Dieks1988,Peres1988,Chefles1998,Peres1998,Sun2001,Rudolph2003,Raynal2005,Raynal2007,Jafarizadeh2008,Sugimoto2010,Waldherr2012,Zhou2012,Bergou2013}. 

High-dimensional quantum states are an important resource for quantum information. In comparison to qubits, the use of qu$d$its, which are states belonging to a $d$-dimensional space, provides access to a larger alphabet and correspondingly higher information rates, and a higher tolerance to noise. The ability to unambiguously discriminate such states is thus of key importance, and successful protocols that accomplish this task will extend the use of these states in quantum information science. Examples of such systems include the time degree of freedom and the spatial light profile, or more specifically the orbital angular momentum degree of freedom, which we use in this work \cite{Mair2001,Leach2010,Agnew2011,Dada2011,Agnew2012,Agnew2013,Donohue2013}. High-dimensional USD is also potentially relevant for pattern recognition in quantum and classical regimes as images contain typically very large numbers of spatial modes and are non-orthogonal to one another \cite{Malik2012}.  

The problem of unambiguous discrimination of qu$d$it states has received a great deal of attention \cite{Herzog2008,Pang2009,Bergou2012,Chen2012,Li2012,Franke-Arnold2012}. USD was first experimentally realised, with a classical light source, to distinguish two non-orthogonal states in the polarisation degree of freedom \cite{Clarke2001}. A subsequent experiment with a similar source extended this to distinguish three states encoded in three-dimensional photon path information \cite{Mohseni2004}. USD has also been performed for two mixed polarisation states using a quantum dot single-photon source \cite{Steudle2011}.

In this work, we discriminate unambiguously between non-orthogonal quantum states encoded in single photons, in dimensions ranging from $d=2$ to $d=14$. While USD theoretically promises the unambiguous discrimination of any set of states, real experimental situations always include error sources, and perfect discrimination in an experimental environment is challenging.  Even with these unavoidable errors, we show that our scheme successfully discriminates between the chosen states and does so with lower error rates than those predicted by MESD. We note that here we implement USD as a sequential measurement of all required detection states. Using instead simultaneous detection, e.g., based on OAM sorter technology \cite{Berkhout2010}, would allow unambiguous discrimination at the single-photon level.

To perfectly distinguish orthogonal states, one requires projections onto the orthogonal state basis, giving $d$ measurement outcomes in a $d$-dimensional space. To implement the USD protocol, which distinguishes non-orthogonal states, one requires the introduction of an additional measurement outcome -- an inconclusive result -- into the procedure, providing $d+1$ measurement possibilities. The increased number of measurement outcomes necessitates the introduction of an ancillary dimension or degree of freedom; orbital angular momentum lends itself well to this treatment as it provides an unlimited supply of additional dimensions. The introduction of the inconclusive result enables the remaining measurement outcomes to be orthogonalised \cite{Neumark1943}. The protocol then provides one of the following: a correct state identification, in which case the state is known with certainty, or an inconclusive result, in which case no information is known about the state.

\begin{figure}
 \centering
 \includegraphics[]{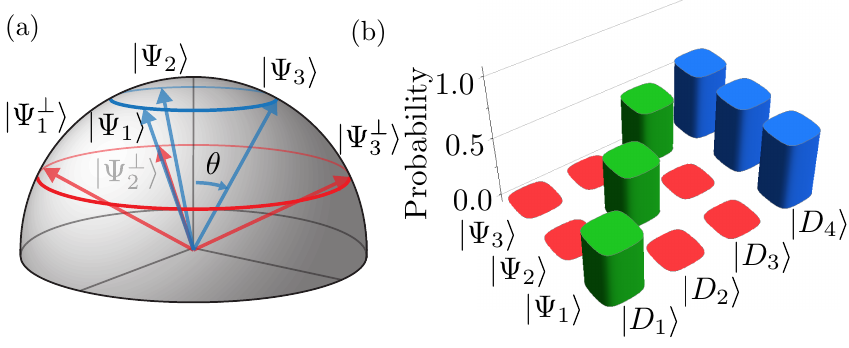}
 \caption{\footnotesize{{\bf Vectors and probabilities in three dimensions.} (a) As the states we consider have real amplitudes, they can be represented on a sphere whose axes are the amplitudes of each basis vector. The vectors we choose to discriminate in dimension $d=3$, $\{\ket{\Psi_i}\}$ with $\theta \approx 33^\circ$, are shown in blue. Vectors perpendicular to each pair ($\{\ket{\Psi^\perp_i}\}$) are shown in red. (b) Theoretically calculated probabilities of discrimination for the vectors shown in (a).}}
 \label{theory}
\end{figure}
In this work, we choose $d$ states in $d$ dimensions that have an equal overlap with each other; these are referred to as equally probable, linearly independent, symmetrical states and, compared to less symmetric states, have a maximal discrimination probability \cite{Chefles1998} \footnote{See Supplementary Materials for further information.}. See Fig.~\ref{theory}(a) for an example in three dimensions. Note that all of these states have only real amplitudes. The overlap between any two states is then a function of the parameter $\theta$, given by
\begin{equation}\label{overlap}
\braket{\Psi_i}{\Psi_j}=\frac{d\,{\rm cos}^2\theta-1}{d-1},
\end{equation}
for $i \neq j$. To ensure positive overlap between the input states, the maximum value of $\theta$ is $\theta_{\rm max}={\rm cos}^{-1}\sqrt{1/d}$ \footnotemark[\value{footnote}].

In the problem of USD, we must establish a set of measurement states $\{\ket{D_i}\}$ to distinguish the set of input states $\{\ket{\Psi_i}\}$. To achieve this, for every state $\ket{\Psi_i}$ we first identify a preliminary measurement state $\ket{\Psi^\perp_i}$; this preliminary state is orthogonal to all other states $\ket{\Psi_j}$ (for $j \neq i$) but has a nonzero overlap with $\ket{\Psi_i}$. Due to this definition, a detection with $\proj{\Psi^\perp_i}$ will unambiguously indicate that the photon was in state $\ket{\Psi_i}$.  These $d$ preliminary measurement states $\{\ket{\Psi^\perp_i}\}$, however, do not generally form an orthonormal basis set.  This can be achieved by extending the preliminary measurement states to an ancillary dimension, followed by normalisation to obtain $d$ measurement states $\{\ket{D_i}\}$.  The basis set is completed by including an additional state $\ket{D_{d+1}}$  orthogonal to all other measurement states, so that the whole $(d+1)$-dimensional basis of measurement states is $\{\ket{D_i}\}$ with $\braket{D_i}{D_j}=\delta_{ij}$.

The probability of obtaining an inconclusive result, $|\braket{\Psi_i}{D_{d+1}}|^2$, and the probability of correctly identifying a state, $|\braket{\Psi_i}{D_i}|^2$, sum to unity as the probability of an error is by definition zero. The probability of an inconclusive result is precisely the overlap between any two input states \cite{Dieks1988,Chefles1998}. Thus using Eq.~(\ref{overlap}), we can write the probabilities of successful identification, erroneous identification, and inconclusive result as
\begin{subequations}\label{p}
\begin{align}
&p_{\rm suc}=\frac{d}{d-1}{\rm sin}^2\theta\\
&p_{\rm err}=0\\
&p_{\rm inc}=\frac{d\,{\rm cos}^2\theta-1}{d-1}.
\end{align}
\end{subequations}
Theoretical predictions of these values for states in three dimensions are shown in Fig.~\ref{theory}(b).

\begin{figure}
 \centering
 \includegraphics[]{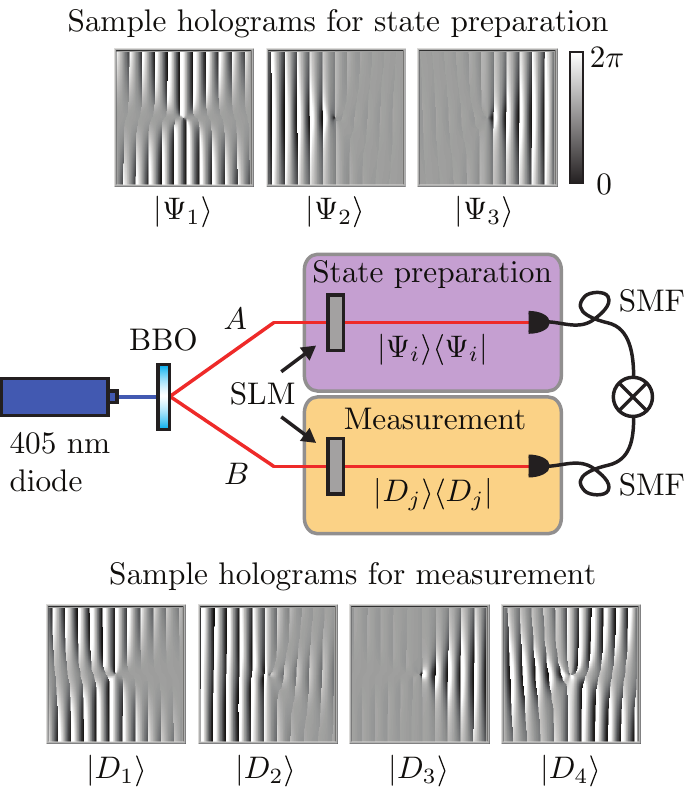}
 \caption{\footnotesize{{\bf Experimental setup.} A pair of entangled photons is produced in a BBO crystal. Arm $A$ is used to prepare a state $\ket{\Psi_i}$, indicated by the purple box; arm $B$ is used to perform a measurement $\ket{D_j}$, indicated by the orange box. Each measurement is accomplished using a spatial light modulator (SLM) and single-mode fibre (SMF). The holograms shown are representative of those used for state preparation and measurement in dimension $d=3$.}}
 \label{setup}
\end{figure}

We use the process outlined above to find the discrimination states for a range of input states in a range of dimensions, and we use them to implement USD as a sequential measurement on orbital angular momentum states. Our experimental procedure is as follows. We produce entangled photons by spontaneous parametric downconversion (SPDC) \cite{Walborn2004} in a 3-mm type-I BBO crystal with a phase mismatch factor of approximately $\phi=-1$. We pump the crystal with a 100-mW laser at 405 nm. In each path, we image the plane of the BBO crystal to a different section of a spatial light modulator (SLM), allowing us to manipulate both the phase and the amplitude of each photon's mode with high fidelity. The simplified experimental setup is shown in Fig.~\ref{setup}.

The photons produced from the BBO crystal are entangled in their orbital angular momentum in the two-photon state $\ket{\psi}=\sum_{\ell=-\infty}^{\infty} c_\ell \ket{\ell}_A \otimes \ket{-\ell}_B$, where $|c_\ell|^2$ is the probability of finding photon $A$ with OAM $\ell \hbar$ and photon $B$ with OAM $-\ell \hbar$ \cite{Torres2003}. The SLM in our experiment performs two functions in regards to this state: first, it allows us to select a range of OAM values and explore a discrete dimension space, and second, it allows us to equalise the probabilities of detection, a process similar to entanglement concentration \cite{Bennett1996}.

The entanglement of the OAM degree of freedom allows the use of remote state preparation \cite{Bennett2001,Liu2007}, which enables us to herald the presence of a range of single-photon states $\ket{\Psi_i}$. These heralded states are prepared by using one half of the SLM in combination with a single-mode fibre. Consequently, the detection of a single photon in the first arm collapses the photon in the other arm into the desired state. The second path is then used to perform the state discrimination measurements $\ket{D_j}$ on the heralded state $\ket{\Psi_i}$, and we measure the coincidences between the two paths.

For a given dimension $d$, we measure all $d+1$ measurement outcomes for each input state $\ket{\Psi_i}$. We use our measurements to calculate a quantity called the quantum contrast, which is defined by the coincidence rates normalised by the singles $Q_{ij}=C_{ij}/(S_{Ai} S_{Bj} t)$; this accounts for any variations in the quantum efficiency of the detection and generation of particular states. Here $C_{ij}$ is the number of coincidence counts defined by an event in both detectors within a time window of $t=25$ ns. The quantities $S_{Ai}$ and $S_{Bj}$ represent the number of counts in path $A$ (heralding the preparation of $\ket{\Psi_i}$) and $B$ (measuring $\ket{D_j}$) respectively. We normalise this quantum contrast into probabilities using $P_{ij}=(Q_{ij}-1)/\sum_j(Q_{ij}-1)$. The $-1$ term accounts for the fact that two independent and uncorrelated sources will have a quantum contrast equal to unity. An integration time of 30 s was used for each measurement, and the maximal coincidence count rate was approximately 350 Hz.

\begin{figure}
 \centering
 \includegraphics[]{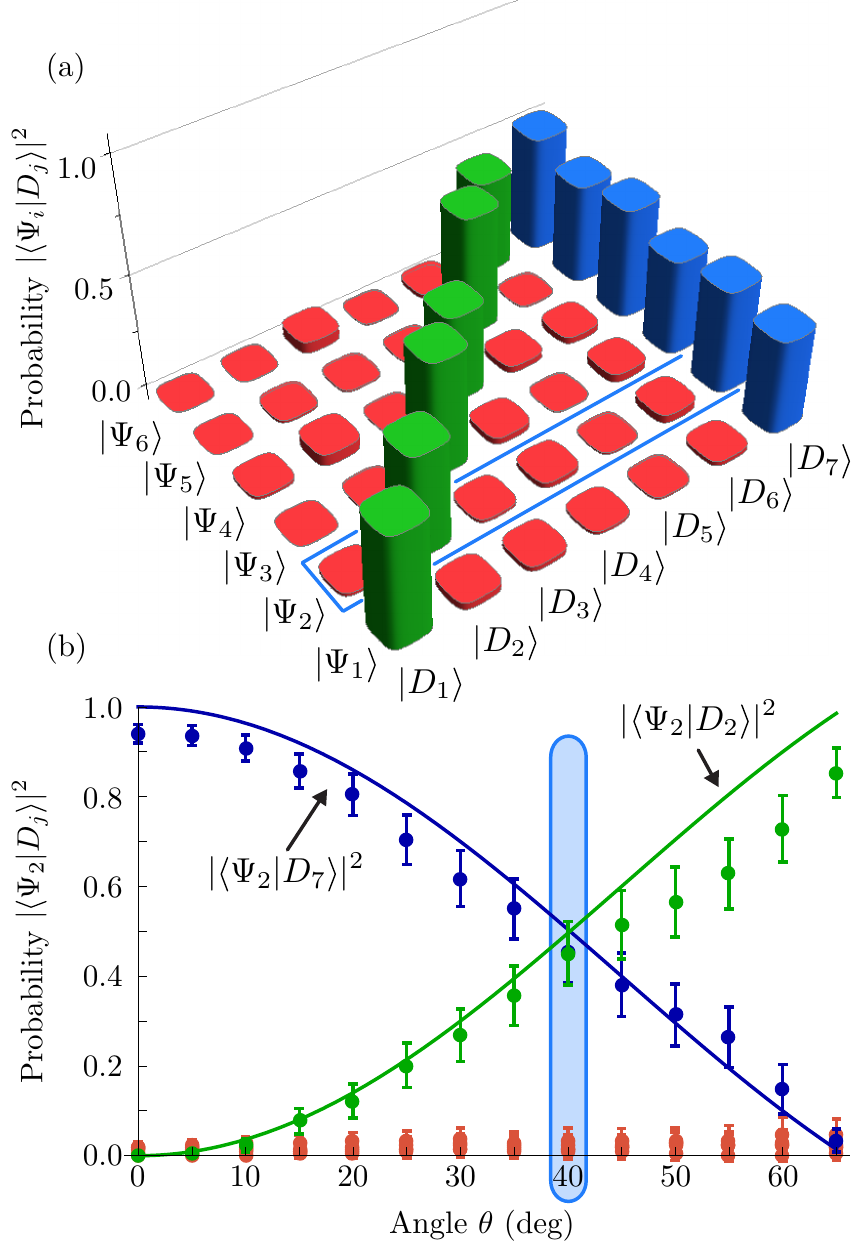}
 \caption{\footnotesize{{\bf Experimental results for dimension $d=6$.} (a) Probabilities of detecting input states $\ket{\Psi_i}$ using detection states $\ket{D_j}$ when $\theta=40^\circ$. (b) Probabilities as a function of $\theta$ of identifying the state $\ket{\Psi_2}$ correctly (green, $\ket{D_2}$), incorrectly (red, $\ket{D_1}, \ket{D_3}, \ket{D_4}, \ket{D_5}, \ket{D_6}$), or inconclusively (blue, $\ket{D_7}$). The points represent experimental data, while the solid lines represent theoretical values calculated using Eq.~(\ref{p}). The points within the shaded area in part (b) correspond to the blue outlined box in part (a). The uncertainties were calculated using Gaussian error propagation, where the measured counts $N$ were assumed to have standard deviation $\sqrt{N}$.}}
 \label{exptdata}
\end{figure}

We have implemented our procedure for unambiguous discrimination of states in high dimensions ranging from $d=2$ to $d=14$ and with varying overlap between the states. In Fig.~\ref{exptdata}, we show the unambiguous discrimination of 6 states in $d=6$ dimensions.

Fig.~\ref{exptdata}(a) shows the results at $\theta=40^\circ$ of measuring all $\{\ket{\Psi_i}\}$ states using all $\{\ket{D_j}\}$ measurements. The green bars denote successful identifications, the red bars denote erroneous identifications, and the blue bars denote inconclusive results. As the probabilities of successful identification greatly exceed the probabilities of erroneous identification, it follows that each input state $\ket{\Psi_i}$ almost always results in either correct detection by $\ket{D_i}$ or the inconclusive outcome $\ket{D_7}$.

Fig.~\ref{exptdata}(b) shows the results of measuring a specific state, in this case $\ket{\Psi_2}$, using all $\{\ket{D_j}\}$ measurements, for a range of angles $\theta$. Each angle corresponds to a different overlap between the $\{\ket{\Psi_i}\}$ states as in Eq.~(\ref{overlap}). An angle of $0^\circ$ corresponds to a complete overlap between the states and hence a completely inconclusive result; the probability for correct identification increases with $\theta$, with in principle perfect identification at $\theta\approx66^\circ$. The solid lines indicate theoretical predictions from Eq.~(\ref{p}); our experimental data is in good agreement with these predictions.

Whilst USD has the theoretical advantage of never misidentifying a state, in practice this is not possible to achieve. In experimental implementations, errors necessarily occur due to finite detector efficiency and errors caused by transformation optics. To evaluate the performance of our measurements, we compare our experimentally recorded errors to those theoretically predicted for the MESD protocol.  A significant advantage is found in the case that the recorded errors for our scheme are smaller than those produced in MESD.

Due to the equal overlap between our input states, the minimum error rate for MESD in $d$ dimensions \cite{Qiu2008} \footnotemark[\value{footnote}] reduces to
\begin{equation}\label{bound}
p_{\rm err} \geq \frac{1}{2}\left(1-\sqrt{1-|\braket{\Psi_i}{\Psi_j}|^2}\right),
\end{equation}
where the overlap $\braket{\Psi_i}{\Psi_j}$ is given by Eq.~(\ref{overlap}). A violation of this inequality indicates that USD provides less ambiguity in state identification than is theoretically possible using MESD.

\begin{figure}
 \centering
 \includegraphics[]{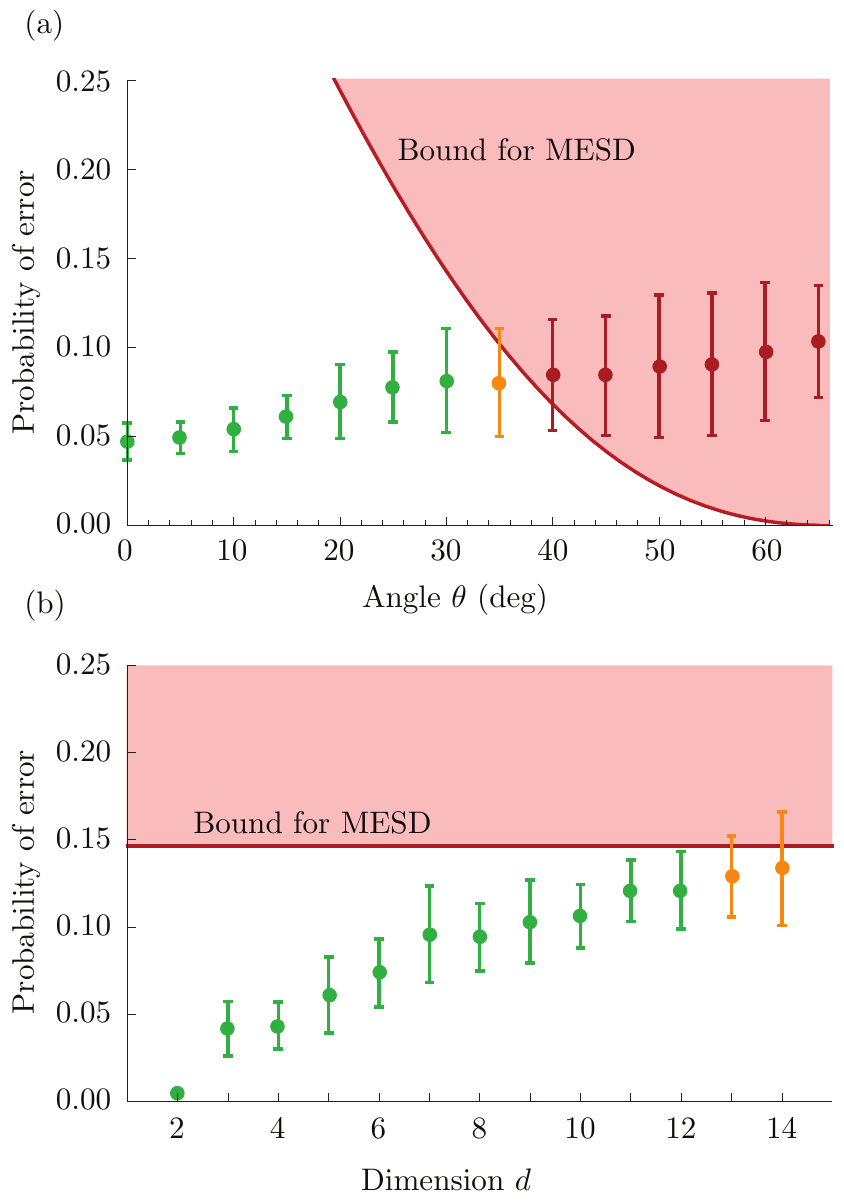}
 \caption{\footnotesize{{\bf Probability of error.} (a) Mean total error rate as a function of angle in dimension $d=6$. (b) Mean total error rate as a function of dimension. Here the angle $\theta$ is chosen individually for each dimension such that the MESD bound is the same in all dimensions. In both plots, the red line indicates the theoretical minimum error rate predicted for MESD. The green points denote error rates at least one standard deviation below this limit, the orange points denote error rates whose uncertainties extend above the limit, and the red points denote error rates above the limit. The uncertainties are the standard deviations associated with the mean values.}}
 \label{errors}
\end{figure}

In Fig.~\ref{errors}, we compare this bound to the mean total error rate observed using our method. To determine our error rate, we first determine the error rate for a single input state $\ket{\Psi_i}$; this is the sum of all possible incorrect state identifications. We then average over all input states $\{\ket{\Psi_i}\}$ to obtain the mean total error rate.

Fig.~\ref{errors}(a) shows the total error rate as a function of angle for the $d=6$ case. The total error rate for angles up to $\theta=30^\circ$ is at least one standard deviation below the MESD bound, demonstrating that our approach is particularly successful for states with large overlap. The total error rate exceeds the MESD bound at higher angles, where the states have lower overlap and are closer to orthogonal. In this case, the bound converges to 0, matching the theoretical prediction for USD. Since the two schemes converge, it is inevitable that the experimentally measured errors exceed the ideal MESD curve at a sufficiently high angle.

Fig.~\ref{errors}(b) shows the total error rate as a function of dimension for a fixed overlap of $1/\sqrt{2}$ between the initial states. We choose a constant overlap so that the MESD bound is equal in all dimensions (in this case, $(1-\sqrt{1/2})/2\approx 0.146$). To achieve the constant overlap, the parameter $\theta$ must change with dimension \footnotemark[\value{footnote}]. The total error rate for dimensions up to $d=12$ is below the MESD bound by at least one standard deviation.

In dimensions $d \geq 13$, the bound for MESD is successfully violated, but by less than one standard deviation. This is due to two main factors. Firstly, for all of these data, the average measured probability of obtaining an error, i.e., measuring a state $\ket{\Psi_i}$ with an incorrect detection state $\ket{D_j}$ ($i \not\in \{j,d+1\}$), is approximately $1\%$. As the dimension increases, so too does the number of opportunities to misidentify a state. Thus the total error grows accordingly, making it increasingly difficult to obtain a low total error. Secondly, due to the limited spiral bandwidth in the downconverted state, the probability amplitudes of the individual OAM modes decrease as $\ell$ increases. This limits the coincidence rate, and thus increases the uncertainty of the measurements, for high dimensions.

We have demonstrated USD via sequential measurements to distinguish $d$ non-orthogonal single-photon states in $d$-dimensional Hilbert spaces. In a modified set-up, our method could be realised as a true POVM experiment in high dimensions. While experimental constraints prevent completely error-free identification, we have shown that, for a range of high-dimensional states, our method still provides a lower error rate than minimum-error state discrimination. With suitable improvements in SLM resolution, spiral bandwidth production, and detector efficiency, this could be increased to even higher dimensions. This method of state discrimination will allow the use of high-dimensional non-orthogonal states in quantum protocols, enabling secure quantum communication with larger alphabets.

We thank Sarah Croke for valuable discussions regarding this work. MA acknowledges financial support from the Natural Sciences and Engineering Research Council of Canada (NSERC).

\section{Supplementary Materials}

\maketitle

\textit{Determining $d$ symmetric states in $d$ dimensions}: The $d$ states in $d$ dimensions that we choose to distinguish are maximally separated when projected onto $d-1$ dimensions. We describe first how to construct the $d-1$ projected vectors $\{\ket{\Psi^\prime_i}\}$. Without loss of generality, the first of these vectors, $\ket{\Psi^\prime_1}$, can be chosen to lie along one axis such that its first component is 1 and its remaining components are 0. We can construct all other vectors from their pairwise overlap, $\braket{\Psi^\prime_i}{\Psi^\prime_j}=-1/(d-1)$ for $i \neq j$, and the normalisation condition, $\braket{\Psi^\prime_i}{\Psi^\prime_i}=1$.  The overlap condition requires that the first component of each remaining vector $\ket{\Psi^\prime_{j\geq 2}}$ must be $-1/(d-1)$. For the second vector we can determine the second component from the normalisation condition and set all following components equal to zero.  The remaining vectors can be iteratively determined in the same way:  for the third vector, the second component is determined from the overlap with the second vector, the third from normalisation, and all following components are zero; and similar for all subsequent vectors.

Once we have obtained these states, we transform them into $d$-dimensional states using
\begin{eqnarray}\label{psii}
\ket{\Psi_i}={\rm sin}\,\theta\,\ket{\Psi^\prime_i} + {\rm cos}\,\theta \,\ket{d}.
\end{eqnarray}
For example, in dimension $d=3$, the states are the lifted trine states
\begin{subequations}
\begin{eqnarray}
\ket{\Psi_1}&=&{\rm sin}\,\theta\,\ket{\ell_1}+{\rm cos}\,\theta\,\ket{\ell_3}\\
\ket{\Psi_2}&=&-\frac{1}{2}{\rm sin}\,\theta\,\ket{\ell_1}+\frac{\sqrt{3}}{2}{\rm sin}\,\theta\,\ket{\ell_2}+{\rm cos}\,\theta\,\ket{\ell_3}\\
\ket{\Psi_3}&=&-\frac{1}{2}{\rm sin}\,\theta\,\ket{\ell_1}-\frac{\sqrt{3}}{2}{\rm sin}\,\theta\,\ket{\ell_2}+{\rm cos}\,\theta\,\ket{\ell_3},
\end{eqnarray}
\end{subequations}
where $\ket{\ell_1}$, $\ket{\ell_2}$, and $\ket{\ell_3}$ are the three chosen OAM basis states.

\vspace{0.5cm}
\textit{Determining discrimination states}: We determine the orthogonal states $\{\ket{\Psi^\perp_i}\}$, with $\braket{\Psi^\perp_i}{\Psi_j} \propto \delta_{ij}$, by taking each $(d-1)$-sized subset of the $\{\ket{\Psi_i}\}$ vectors and applying the Gram-Schmidt algorithm to find a vector orthogonal to this subset.

We then transform the set $\{\ket{\Psi^\perp_i}\}$ to an orthonormal basis set $\{\ket{D_i}\}$ by extension to an ancillary dimension followed by normalisation. For this, we make use of the fact that due to the inherent symmetry, the inner product of any two of the $\{\ket{\Psi^\perp_i}\}$ states, $\braket{\Psi^\perp_i}{\Psi^\perp_j}$, $i \neq j$, is the same. As a result, we obtain
\begin{equation}
\ket{D_i}=\ket{\Psi^\perp_i}+\sqrt{-\braket{\Psi^\perp_1}{\Psi^\perp_2}}\ket{d+1}.
\end{equation}
Finally, we identify the inconclusive measurement state $\ket{D_{d+1}}$ such that $\braket{D_i}{D_{d+1}}=0$, resulting in a complete basis in $d+1$ dimensions, again using the Gram-Schmidt algorithm. 

\vspace{0.5cm}
\textit{Transforming $\ket{\Psi^\perp_i}$ to $\ket{D_i}$}: Here we illustrate the calculation for $d=3$, but it functions similarly in higher dimensions. In order to orthogonalise our three 3-dimensional measurement states
\begin{subequations}
\begin{eqnarray}
\ket{\Psi^\perp_1}&=&\sqrt{3}{\rm cos}\,\theta\,{\rm sin}\,\theta\,\ket{\ell_1}+\frac{\sqrt{3}}{2}{\rm sin}^2\theta\,\ket{\ell_3}\\
\ket{\Psi^\perp_2}&=&-\frac{\sqrt{3}}{2}{\rm cos}\,\theta\,{\rm sin}\,\theta\,\ket{\ell_1}+\frac{3}{2}{\rm cos}\,\theta\,{\rm sin}\,\theta\,\ket{\ell_2}\notag\\&+&\frac{\sqrt{3}}{2}{\rm sin}^2\theta\,\ket{\ell_3}\\
\ket{\Psi^\perp_3}&=&-\frac{\sqrt{3}}{2}{\rm cos}\,\theta\,{\rm sin}\,\theta\,\ket{\ell_1}-\frac{3}{2}{\rm cos}\,\theta\,{\rm sin}\,\theta\,\ket{\ell_2}\notag\\&+&\frac{\sqrt{3}}{2}{\rm sin}^2\theta\,\ket{\ell_3}
\end{eqnarray}
\end{subequations}
into three 4-dimensional measurement states $\{\ket{D_1},\ket{D_2},\ket{D_3}\}$, we need only to add an arbitrary fourth component to each vector such that
\begin{equation}
\ket{D_j}=\ket{\Psi^\perp_j}+(a_j+ib_j)\ket{d+1},
\end{equation}
where $a_i$ and $b_i$ are real numbers. For orthogonality, these states must satisfy
\begin{equation}
\braket{D_i}{D_j}=C\delta_{ij},
\end{equation}
where $C$ is some constant since the vectors are as yet unnormalised.

By examining the inner product $\braket{D_1}{D_2}$, we obtain
\begin{equation}
\braket{D_1}{D_2}=0=\braket{\Psi^\perp_1}{\Psi^\perp_2} + a_1 a_2 + i a_1 b_2 - i b_1 a_2 + b_1 b_2.
\end{equation}
The real and imaginary parts then independently need to be equal to zero:
\begin{eqnarray}
\braket{\Psi^\perp_1}{\Psi^\perp_2} + a_1 a_2 + b_1 b_2 &=& 0\label{d1d2}\\
a_1 b_2 - b_1 a_2 &=& 0.
\end{eqnarray}
From the inner products $\braket{D_2}{D_3}$ and $\braket{D_1}{D_3}$, we obtain similar equations. These six equations are solved simultaneously by defining the coefficients $a_i$ and $b_i$ as
\begin{subequations}
\begin{eqnarray}
a_2&=&a_1 \frac{\braket{\Psi^\perp_2}{\Psi^\perp_3}}{\braket{\Psi^\perp_1}{\Psi^\perp_3}}\\
b_2&=&b_1 \frac{\braket{\Psi^\perp_2}{\Psi^\perp_3}}{\braket{\Psi^\perp_1}{\Psi^\perp_3}}\\
a_3&=&a_1 \frac{\braket{\Psi^\perp_2}{\Psi^\perp_3}}{\braket{\Psi^\perp_1}{\Psi^\perp_2}}\\
b_3&=&b_1 \frac{\braket{\Psi^\perp_2}{\Psi^\perp_3}}{\braket{\Psi^\perp_1}{\Psi^\perp_2}}.
\end{eqnarray}
\end{subequations}
By substituting these values back into Eq.~(\ref{d1d2}), we obtain
\begin{eqnarray}
0&=&\braket{\Psi^\perp_1}{\Psi^\perp_2}+\frac{\braket{\Psi^\perp_2}{\Psi^\perp_3}}{\braket{\Psi^\perp_1}{\Psi^\perp_3}}(a_1^2+b_1^2)\\
a_1^2+b_1^2&=&-\frac{\braket{\Psi^\perp_1}{\Psi^\perp_2}\braket{\Psi^\perp_1}{\Psi^\perp_3}}{\braket{\Psi^\perp_2}{\Psi^\perp_3}}.
\end{eqnarray}
However, in our particular case, we know that the overlap between each pair of vectors in the set $\{\ket{\Psi^\perp_i}\}$ is equal; as a result, this can be reduced to
\begin{equation}
a_1^2+b_1^2=-\braket{\Psi^\perp_1}{\Psi^\perp_2}.
\end{equation}
Recall that we defined $a_i,b_i \in \mathbb{R}$; thus $a_1^2+b_1^2 \geq 0$ and we find that the states $\{\ket{D_i}\}$ as defined above can only exist if
\begin{equation}
\braket{\Psi^\perp_1}{\Psi^\perp_2} \leq 0.
\end{equation}

In order to form our $\ket{D_i}$ states, we choose $b_1=0$ for simplicity and thus $a_1=\sqrt{-\braket{\Psi^\perp_1}{\Psi^\perp_2}}$ and our discrimination states become
\begin{equation}
\ket{D_i}=\ket{\Psi^\perp_i}+\sqrt{-\braket{\Psi^\perp_1}{\Psi^\perp_2}}\ket{d+1}.
\end{equation}
The normalised form of these states is
\begin{equation}
\ket{D_i}=\frac{\ket{\Psi^\perp_i}+\sqrt{-\braket{\Psi^\perp_1}{\Psi^\perp_2}}\ket{d+1}}{(\braket{\Psi^\perp_i}{\Psi^\perp_i}-\braket{\Psi^\perp_1}{\Psi^\perp_2})^2}.
\end{equation}
The inconclusive result $\ket{D_4}$ can then be found using the Gram-Schmidt algorithm on the first three $\{\ket{D_i}\}$ vectors.

For our example of dimension $d=3$, we end up with the following discrimination states:
\begin{widetext}
\begin{subequations}
\begin{eqnarray}
\ket{D_1}&=&\sqrt{\frac{2}{6}}\ket{\ell_1}+\frac{1}{\sqrt{6}}{\rm tan}\,\theta\ket{\ell_3}+\sqrt{\frac{3\,{\rm cos}^2\theta-1}{6}}{\rm sec}\,\theta\ket{\ell_4}\\
\ket{D_2}&=&-\frac{1}{\sqrt{6}}\ket{\ell_1}+\frac{1}{\sqrt{2}}\ket{\ell_2}+\frac{1}{\sqrt{6}}{\rm tan}\,\theta\,\ket{\ell_3}+\sqrt{\frac{3\,{\rm cos}^2\theta-1}{6}}{\rm sec}\,\theta\ket{\ell_4}\\
\ket{D_3}&=&-\frac{1}{\sqrt{6}}\ket{\ell_1}-\frac{1}{\sqrt{2}}\ket{\ell_2}+\frac{1}{\sqrt{6}}{\rm tan}\,\theta\,\ket{\ell_3}+\sqrt{\frac{3\,{\rm cos}^2\theta-1}{6}}{\rm sec}\,\theta\ket{\ell_4}\\
\ket{D_4}&=&-\sqrt{\frac{3\,{\rm cos}^2\theta-1}{2}}{\rm sec}\,\theta\ket{\ell_3}+\frac{1}{\sqrt{2}}{\rm tan}\,\theta\ket{\ell_4}.
\end{eqnarray}
\end{subequations}
\end{widetext}

\vspace{0.5cm}
\textit{Angle calculation for fixed overlap}: From the definition of the vector  $\ket{\Psi_i}$ in  Eq.~(\ref{psii}), we find that the overlap is
\begin{equation}
\braket{\Psi_i}{\Psi_j}={\rm sin}^2\theta\braket{\Psi^\prime_i}{\Psi^\prime_j}+{\rm cos}^2\theta,
\end{equation}
where we have used that all vectors $\ket{\Psi_i}$ are orthogonal to $\ket{d}$.  Using furthermore that all vectors have the same overlap $\braket{\Psi^\prime_i}{\Psi^\prime_j}=-1/(d-1)$ we find
\begin{eqnarray}
\braket{\Psi_i}{\Psi_j}&=&{\rm sin}^2\theta\left(-\frac{1}{d-1}\right)+{\rm cos}^2\theta \nonumber \\
&=&\frac{d \, \cos^2\theta-1}{d-1}. \label{overlap2}
\end{eqnarray}

In order to obtain states with equal overlap but defined in different dimensions, as we have chosen for Fig.~4(b), we can solve the above equation for $\theta$,
\begin{equation}
\theta={\rm cos}^{-1}\sqrt{\frac{1}{d}(1+(d-1) \braket{\Psi_i}{\Psi_j})}.
\end{equation}

\vspace{0.5cm}
\textit{MESD bound in $d$ dimensions}: As shown in Ref.~\cite{Qiu2008}, the error obtained using MESD to distinguish $d$ states in $d$ dimensions satisfies the inequality
\begin{equation}
p_{\rm err} \geq \frac{1}{2}\left( 1 - \frac{1}{d-1} \sum_{i=1}^d \sum_{j=1}^{i-1} {\rm Tr} \left| \eta_i \rho_i - \eta_j \rho_j \right| \right),\label{MESDbound}
\end{equation}
where $\eta_i$ is the \textit{a priori} probability of generating the state $\rho_i$ and $|X|=\sqrt{X^\dagger X}$.

This expression can be simplified somewhat in our case. Firstly, our \textit{a priori} probabilities $\eta_i$ are all equal to $1/d$ so that
\begin{equation}
{\rm Tr} \left| \eta_i \rho_i - \eta_j \rho_j \right|=\frac{1}{d}{\rm Tr} \left| \rho_i - \rho_j \right|.
\end{equation}
Secondly, we use only pure states, so that \cite{NielsenChuang}
\begin{equation}
{\rm Tr} \left| \rho_i - \rho_j \right|=2\sqrt{1-|\braket{\Psi_i}{\Psi_j}|^2}.
\end{equation}
Then Eq.~(\ref{MESDbound}) becomes
\begin{equation}
p_{\rm err} \geq \frac{1}{2}\left( 1-\frac{1}{d-1}\sum_{i=1}^d \sum_{j=1}^{i-1} \frac{2}{d} \sqrt{1-|\braket{\Psi_i}{\Psi_j}|^2} \right).
\end{equation}
Since all initial states $\{\ket{\Psi_i}\}$ for a particular angle and dimension have a known equal overlap with one another, the term inside the sum is a constant and can be factored out so that
\begin{equation}
p_{\rm err} \geq \frac{1}{2}\left( 1-\frac{1}{d-1} \frac{2}{d} \sqrt{1-|\braket{\Psi_i}{\Psi_j}|^2}\sum_{i=1}^d \sum_{j=1}^{i-1}1 \right).
\end{equation}
By evaluating the sum as $\sum_{i=1}^d \sum_{j=1}^{i-1} 1 = (d^2-d)/2$, we obtain
\begin{equation}
p_{\rm err} \geq \frac{1}{2}\left(1-\sqrt{1-|\braket{\Psi_i}{\Psi_j}|^2}\right).
\end{equation}
As the overlap is defined by Eq.~(\ref{overlap2}), we finally find the MESD error bound to be
\begin{equation}
p_{\rm err} \geq \frac{1}{2}\left[1-\sqrt{1-\left( \frac{d \, \cos^2\theta-1}{d-1} \right)^2}\right].
\end{equation}
For states that overlap by $\braket{\Psi_i}{\Psi_j}=1/\sqrt{2}$, as in Fig.~4(b), this evaluates to $p_{\rm err} \geq \frac{1}{2}\left( 1-\sqrt{\frac{1}{2}}\right)\approx 0.146.$

\vspace{0.5cm}
\textit{OAM values}: For dimension $d$, we require $d+1$ OAM values: $d$ OAM values to form a basis for our states, and one additional OAM value to facilitate our discrimination measurements. As the probability of production of an OAM value decreases in absolute value, it is advantageous to use OAM values closest to zero to obtain greatest signal. The chosen OAM values for several dimensions are shown in Table \ref{table}.

\begin{table}[h]
\begin{tabular}{ | c | c | c | }
  \hline
  Dimension $d$ & $\ell$-values for states & Ancillary $\ell$-value \\ \hline
  2 & 0,1 & -1 \\
  3 & -1,0,1 & -2 \\
  4 & -1,0,1,2 & -2 \\
  5 & -2,-1,0,1,2 & -3 \\
  \hline
\end{tabular}
\caption{OAM values.}
\label{table}
\end{table}

\end{document}